# Highly Cited Papers of Ukrainian Scientists Written in Collaboration: A Bibliometric Analysis (2011-2015)

Serhii Nazarovets, Kyiv National University of Culture and Arts,
14 Chyhorina Str., Kyiv, Ukraine 01042
E-mail: serhii.nazarovets@gmail.com

**Abstract**

The paper presents the results of the study of international and national cooperation of Ukrainian scientists from different scientific fields using citation analysis data from Scopus in the period of 2011-2015. The results show that during the period under study, the number of documents of highly cited Ukrainian scientists that have received enough citations to be included to the top 1%, 5% and 10% most cited documents in the world, evidenced an increase and were significantly different in different subjects areas. Papers written by a group of co-authors predominate among highly cited documents of Ukrainian scientists. Consequently, international cooperation plays an important role in Ukrainian scientist's research visibility and impact. At the same time, up to 16%-27% highly cited articles of Ukrainian scientists have been written without partnering with foreign colleagues, that means a significant part of important scientific research in Ukraine are carried out by oneself. Therefore, for generating an optimal policy of science development in Ukraine it is important to provide a balanced view of the expectations of the results of the international cooperation of Ukrainian scientists and achieving a required balance between the inter-country and international collaboration.

## 1. Introduction

After the collapse of the Soviet Union in 1991, Ukraine declared independence and faced a number of issues typical for "young democracies" including slow transition from a centrally planned economy to a market economy, a sharp reduction of science funding, brain drain, growth of corruption in scientific and educational spheres, inefficient use of budget funds. The model of science functioning, inherited from the Soviet Union, was open to radical reforms required for the modernization of science and education industry, so a long process of stagnation of thought in Ukraine began, to it by a certain voluntary self-isolation and the gradual disappearance from the world's scientific map. Despite a significant decrease in the number of high-quality research, relevant international standards and the number of scientists, research institutions in the country has not changed significantly, and therefore, even those the meager state funding allocated for the national scientific product sprayed, and Ukrainian society and the state does did not receive an adequate benefit from them.

For example, as of 2016, 43 titles of scientific journals published in Ukraine were indexed in Scopus, and only 15 titles - in the Journal Citation Reports. The web portal "Scientific Periodicals of Ukraine" managed by Vernadsky National Library of Ukraine with the aim of providing electronic copies of Ukraine scientific journals on the Internet, contains 2517 titles (as of December 2016) and by some estimates, in fact, the number is over 3000. Most Ukrainian journals have not clearly established scientific subjects and are not included in any international abstracting databases, have a very limited readership, and therefore, they are ineffective as a tool for exchange of scientific information [6]. Therefore, the majority of Ukrainian scientists tend to publish their best works in foreign journals, and the quality of Ukrainian peer-reviewed journals is reduced to the regional level.

Updating science policy management cannot be met by means of a mere cosmetic improvement and will probably require deeper reforms. Over the last ten years, the active participation of Ukrainian scientists in international collaborations stimulated the development of Ukraine's science. Such an active integration, according to officials, should improve the visibility of Ukrainian scientists, reduce the cost of scientific research and, thereby, open up new professional opportunities for numerous researchers. The annexation of Crimea by the Russian Federation in 2014 and the subsequent Russian armed aggression, which is still happening in the East of Ukraine, slowed down the reforms and modernization of the Ukrainian science management system again. However, even in difficult conditions of wartime, it was possible to achieve certain success [8]. In particular, as a consequence of the signed agreement between Ukraine and the European Union, Ukrainian scientists and institutions obtained the ability to fully participate in events funded within the framework of Horizon 2020, the flagship funding program for research and innovation of the European Union. Accordingly, the Ukrainian science managers hope that the successful future integration into the European scientific space will have positive consequences for the modernization of the scientific system of Ukraine [7].

## 2. Literature review

Scientific research collaboration facilitates producing of the new knowledge and expands the perspectives for further use of research results. Cooperation between researchers boosts mutual enrichment by scientific ideas, effective use of skills, competencies, and resources [5]. As a rule, it is considered that international scientific cooperation should be encouraged because it positively correlates with the quality of research and the increase in research activity. But such benefits are not self-evident. The geopolitical situation, cultural relations, language are the determining factors for formation advantages of scientific cooperation and further influence of documents created as a result of such cooperation [9].

International co-authorship, as a rule, produces documents with higher citation rates than in-country work. However, the impact of international cooperation to the citation of national authors varies greatly by country, and there are also noticeable differences between the scientific branches within a given country [4]. The number of citations of scientific paper depends not only on the fact of cooperation but, of course, on who of the authors participates in the work on it. In addition, the high internal national saturation of scientific discipline by documents and researchers sometimes leads to a kind of isolation and inhibits further cooperation [3, 10].

Consequently, in conditions of limited resources, university and research institutions administrators in developing countries need to remember, that scientific cooperation does not always lead to an increase in the publication productivity of scientists. Publication activity is strongly associated with various professional factors, but there is no direct evidence of its dependence on scientific cooperation. Usually, the scientists view joint research projects as a way of obtaining professional opportunities, or some kind of external rewards, rather than as a means to gain new knowledge and academic recognition [11].

Thus, a participation of Ukrainian scientists in international collaborations may not have predictable positive results for the Ukraine science, but rather divert the attention of managers from other promising opportunities. Therefore, the purpose of this study was to determine the specific features of the cooperation of Ukrainian scientists in all academic fields on the basis of an analysis of the citation of their documents indexed in Scopus for the period 2011-2015, and to establish, whether international cooperation is indeed of such exceptional importance for the visibility and impact of research results of Ukrainian scientists in comparison with individual works, or publications written without international participation.

## 3. Data and methodology

We used integrated modular platform SciVal, which analyses the activities of research organizations based on Scopus data. The Scopus database covers all subject areas, has quality selection criteria for indexed publications, allows tracking citation of scientific papers and calculates a number of scientometric indicators for scientists, institutions, and peer-reviewed journals. A high number of references to publications in other scientific papers may indicate that the cited documents were sufficiently influential and did not remain unnoticed in scientist's communities [2]. For the study of the correlation between international cooperation of Ukrainian scientists and the writing of influential scientific publications, we retrieved the documents of Ukrainian scientists for the period 2011-2015, which have reached the top 1%, 5% and 10% of the most cited papers in respective subject areas.

The sample of the study consists of documents of all types that are indexed in Scopus in the period 2011-2015, in which at least one of the authors indicated as an employee of a Ukrainian institution. The data was collected using the search function AFFILCOUNTRY (Ukraine) and exported on December 11, 2016. A certain number of documents probably were not included to the search results due to error in the database or lack of information about the author's affiliation. Most of the selected publications are written by two or more authors from different countries, but for the research needs, if at least one of the authors affiliated his works with Ukrainian institution, the publication is considered "Ukrainian".

All identified documents of Ukrainian scientists were divided into groups, according to the number of authors of the documents: more than 5 co-authors (Multi5 +), 3-5 co-authors (Group 3-5), two co-authors (Double) and documents of single authors (Single Author). In addition, there are two separate groups: the documents written by co-authors only from Ukrainian institutions (National), and presented by scientists working in the same Ukrainian institution (Institutional).

The data on document type were extracted and the dynamics of their distribution over the years for all documents from the sample was determined. And we found the journal subject area for each publication, according to the classification scheme used in Scopus. We have compared the number of highly cited journal publications in each discipline, both for all documents of Ukrainian authors and only for those that were written only in national collaborations. Also, we identified a country (or countries) of origin for each journal and the number of open journals.

## 4. Results and discussion

We found out 285 documents among the world`s top 1% most cited papers (among them 214 journal articles, 44 conference materials, 18 book series, 9 books); 1342 publications of the top 5% most cited documents (including 980 journal articles, 240 conference materials, 82 book series, 39 books, 1 industry publication); 2588 publications of the top 10% most cited documents in their field (including 1888 journal articles, 507 conference proceedings, 122 book series, 70 books, 1 industry publication) for the period of 2011-2015, in which at least one of the authors worked in a Ukrainian institution (Figure 1). A small increase in the number of the most cited papers of Ukrainian scientists noticed for period of five years.

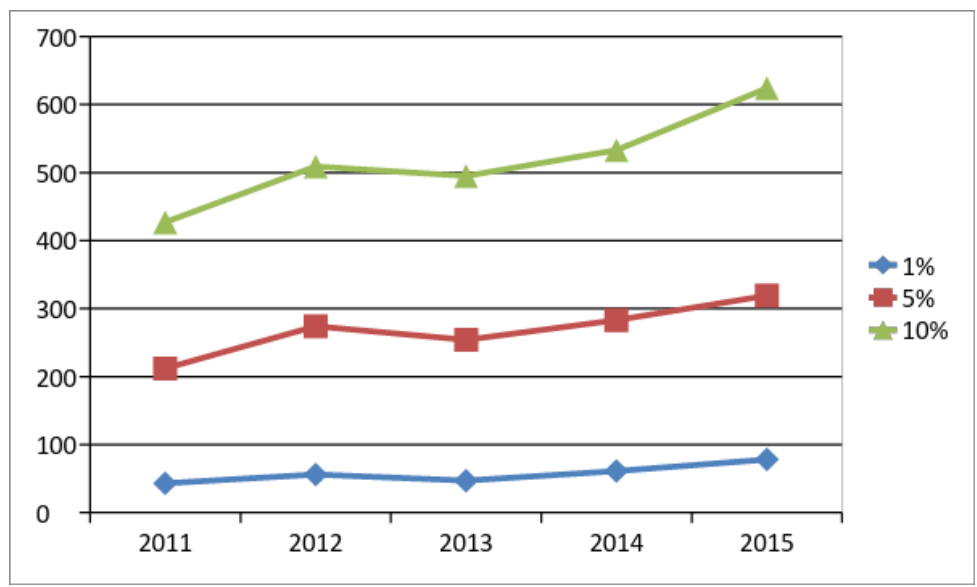

**Figure 1.** The distribution of Ukrainian scientists documents of the top in 1%, 5% and 10% most cited papers in 2011-2015.

The 214 journal articles of the top 1% most cited documents were published in the 103 journals (2.08 articles per journal). The 980 journal articles of the top 5% most cited documents were published in the 417 journals (2.35 articles per journal). And the 1888 journal articles of the top 10% most cited documents were published in the 692 journals (2.7 articles per journal).

Most of the cited articles were published in non-Ukrainian scientific journals. Only 6 articles in the top 1% most cited documents were published in 2 Ukrainian journals. In fact, only 36 articles of the top 5% most cited documents were published in 8 Ukrainian journals. And 59 articles among the articles of the top 10% most cited documents were published in 9 Ukrainian journals.

The papers of Ukrainian scientists of the top 1% most cited documents were published in 13 open journals (13% of all journal titles), papers from top 5% of the most cited documents were published in 57 open journals (14%), papers from the top 10% most of the cited documents were published in 82 open journals (12%). There is not enough collected data to make conclusions about the impact of the openness of a journal on the publication strategy of Ukrainian scientists and about they supporting the Open Access Initiative.

We divided documents of Ukrainian scientists of the top 1%, 5% and 10% most cited papers into 4 groups, according to the number of their authors. The documents which have more than 5 co-authors, were included in the group "Multi5 +", the documents of 3-5 co-authors - the "Group 3-5", the documents of two co-authors - the "Double", and documents of individual authors were included in the group "Single Author". Also, we have identified groups of documents written by co-author scientists only from Ukrainian institutions (National) and documents written by co-authors of scientists working in the same Ukrainian institution (Institutional) (Tables 1-3).

**Table 1.** The types of collaborations of Ukrainian scientists documents of the top 1% most cited papers by Scopus for the period 2011-2015.

| Collaboration type | Description | Number of documents | % |
|---|---|---|---|
| **Multi5+** | Collaboration involving more than 5 researchers | 184 | 64% |
| **Group 3-5** | Collaboration with more than 3-5 researchers | 73 | 26% |
| **Double** | Collaboration involving 2 researchers | 17 | 6% |
| **Single Author** | Individual publications | 11 | 4% |
| **National** | Collaboration involving only employees of Ukrainian institutions | 47 | 16% |
| **Institutional** | Collaboration involving only employees from one Ukrainian institution | 29 | 10% |

**Table 2.** The types of collaborations of Ukrainian scientists documents of the top 5% most cited papers by Scopus for the period 2011-2015.

| Collaboration type | Description | Number of documents | % |
|---|---|---|---|
| **Multi5+** | Collaboration involving more than 5 researchers | 706 | 52% |
| **Group 3-5** | Collaboration with more than 3-5 researchers | 452 | 34% |
| **Double** | Collaboration involving 2 researchers | 116 | 9% |
| **Single Author** | Individual publications | 68 | 5% |
| **National** | Collaboration involving only employees of Ukrainian institutions | 310 | 23% |
| **Institutional** | Collaboration involving only employees from one Ukrainian institution | 210 | 16% |

**Table 3.** The types of collaborations of Ukrainian scientists documents of the top 10% most cited papers by Scopus for the period 2011-2015.

| Collaboration type | Description | Number of documents | % |
|---|---|---|---|
| **Multi5+** | Collaboration involving more than 5 researchers | 1236 | 48% |
| **Group 3-5** | Collaboration with more than 3-5 researchers | 939 | 36% |
| **Double** | Collaboration involving 2 researchers | 252 | 10% |
| **Single Author** | Individual publications | 161 | 6% |
| **National** | Collaboration involving only employees of Ukrainian institutions | 691 | 27% |
| **Institutional** | Collaboration involving only employees from one Ukrainian institution | 356 | 14% |

It was quite expected that more publications among multi-authorship highly cited publications of Ukrainian scientists more than 5 co-authors. At the same time, in the group with the smallest percentage of the most cited documents, we found the largest number of works written in major international collaborations.

The number of individual publications is relatively small and constitutes a small fraction of all the highly cited papers of Ukrainian scientists. However, they make up 4-6% of the total number of highly cited papers of Ukrainian scientists, depending on the percentage of the most cited documents. This shows that even in the current unfavorable conditions for the development of science in Ukraine, individual scientists can successfully carry out influential scientific research.

Among the Ukrainian authors of the top 1% most cited documents in 2011-2015 according to Scopus, only 3 authors have 2 affiliations (representing Ukrainian and foreign institutions). While the rest of the authors work only for one Ukrainian scientific institution (Table 4).

**Table 4.** The top-10 of the Ukrainian authors of the most cited documents of the top 10% most cited papers in 2011-2015 according to Scopus.

| № | Authors | Scopus affiliation names | Country |
|---|---|---|---|
| 1. | Lushchak, V.I. | Precarpathian University | Ukraine |
| 2. | Kordyuk, A.A. | G. V. Kurdyumov Institute for Metal Physics of National Academy of Sciences of Ukraine | Ukraine |
| 3. | Muzyka, K. | Kharkiv National University of Radio Electronics | Ukraine |
| 4. | Zaslavskii, O.B. | Kharkiv National University | Ukraine |
| 5. | Omel'chenko, O.E. | Institute of Mathematics of National Academy of Sciences in Ukraine, Weierstrass Institute for Applied Analysis and Stochastics | Ukraine, Germany |
| 6. | Savelyev, A. | National Academy of Sciences in Ukraine, The University of North Carolina at Chapel Hill | Ukraine, USA |
| 7. | Zaslavskii, O.B. | Kharkiv National University | Ukraine |
| 8. | Pogrebnjak, A.D. | Sumskij Derzavnij Universitet | Ukraine |
| 9. | Wasser, S.P. | National Academy of Sciences in Ukraine, University of Haifa | Ukraine, Israel |
| 10. | Bekshaev, A.Y. | Odessa National University | Ukraine |

The presence of an internal national scientific potential becomes even more evident if one considers that as many as 27% of the most cited papers, found in 10% of the most cited documents, were written only by scientists working in Ukrainian institutions.

To identify the subject of collaborations in different subject areas, for each journal publications of the top 1%, 5% and 10% most cited documents, the journal subject area was determined according to the classification scheme used in Scopus (Table 5). According to the number of quoted documents of Ukrainian scientists, the Physics and Astronomy has higher positions, then we see Materials Science, Engineering, Chemistry, Medicine and Mathematics (Figure 2).

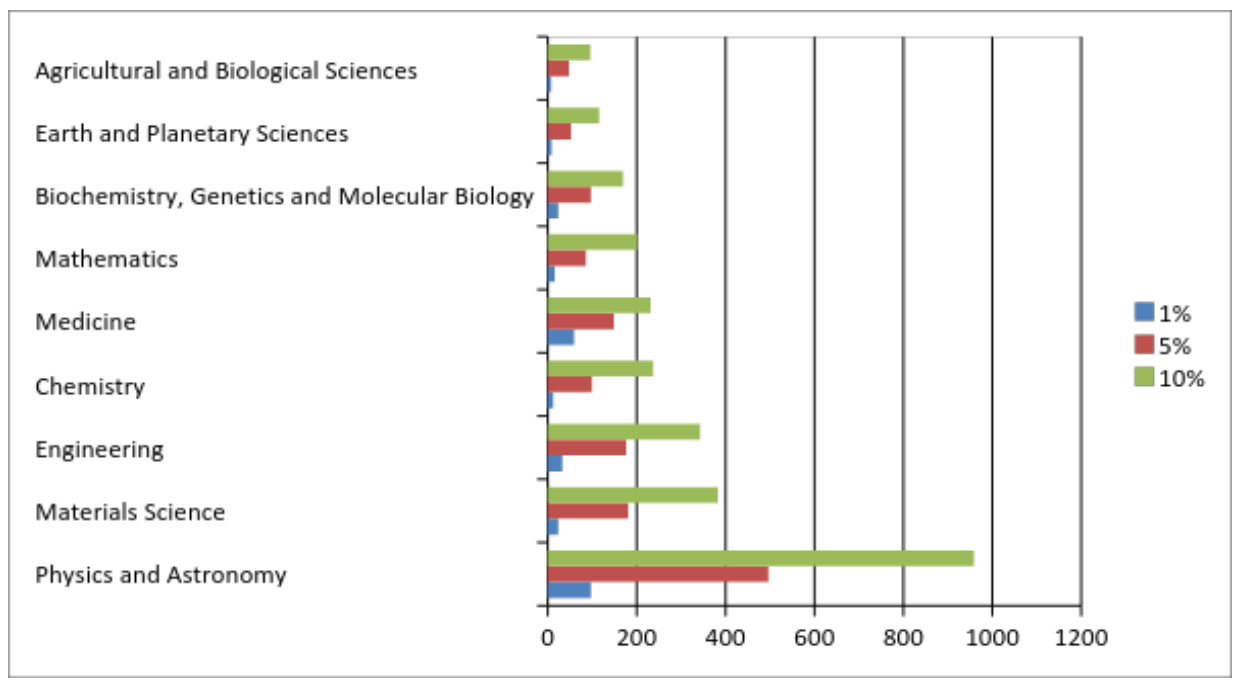

**Figure 2.** The top-10 subject areas by the number of journal publications of Ukrainian scientists of the top 1%, 5% and 10% most cited papers in 2011-2015.

The Subject Area Categories of individual journal publications of Ukrainian scientists and publications written in collaborations only with employees of Ukrainian institutions, which were included in world`s top 1%, 5% and 10% of the most cited papers for the period 2011-2015, similar to the general thematic distribution, but it also has their own characteristics (Table 6). In this distribution, the Physics and Astronomy, Engineering, Materials Science, Mathematics are on leading positions again. However, the influence of international collaborations turned out to be quite appreciable for the Medicine publications (Figure 3).

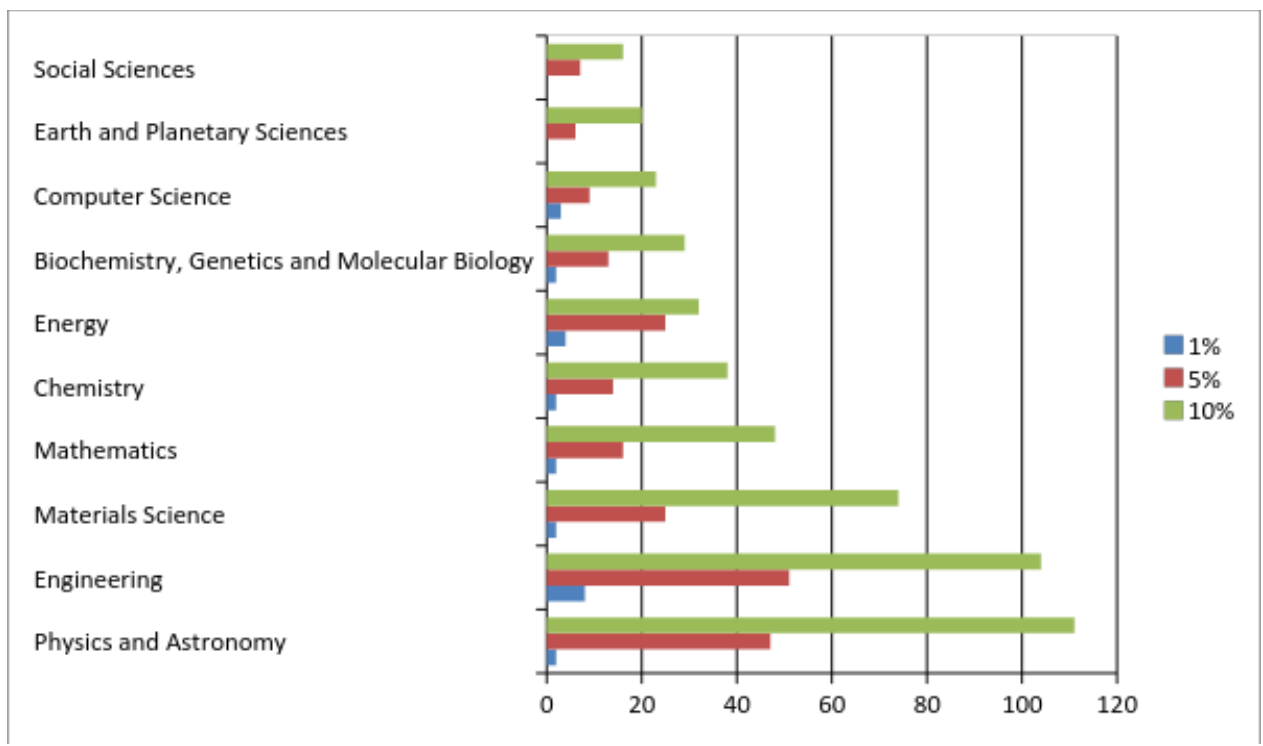

**Figure 3.** The top-10 subject areas of individual journal publications of the Ukrainian scientists and the journal publications written in collaborations only with employees of Ukrainian institutions, which are among world`s top 1%, 5% and 10% of the most cited papers for the period 2011-2015.

In addition, we found out that there are no highly cited exclusively national publications in the areas of Dentistry, Management and Accounting, Nursing, Veterinary in Ukraine, which may indicate a lack of funding and infrastructures for conducting qualitative research.

**5. Conclusion**

The objective of this study was to identify the specifics of the cooperation of Ukrainian scientists in various scientific fields by analysing highly cited papers for the period 2011-2015 according to Scopus and the role international cooperation plays in making results of the research of Ukrainian scientists visible and influential. During the period under review, the number of highly cited documents of Ukrainian scientists of the top 1%, 5% and 10% most cited documents grew steadily and differed significantly in different scientific fields. The Ukrainian scientists in Physics and Astronomy were particularly productive (959 documents). Next in productivity, with a significant margin, were publications of the scientists in Materials Science (383 documents) and Engineering (343 documents).

Highly cited articles of Ukrainian authors appeared in Ukrainian scientific journals extremely rarely. In this respect, we can state that, despite the huge number of scientific journal titles published in Ukraine, these journals do not perform their role as a reliable channel of scientific communication properly.

Besides, not only journal articles but also books and other types of documents play an important role in the process of scientific communication in the social and human sciences. In addition, much of the research in the humanities reaches primarily national audience. Consequently, similar studies appear exclusively in regional journals and monographs of local publishing houses, which may not be visible to the global academic community. However, despite these limitations, problems with communication channels in the Ukrainian social and, especially, humanities sciences have been known for a long time, and the results of this study confirm previous findings [6].

It should be noted that in the search for publisher countries, translation editions were not taken into account - those Ukrainian journals that publish articles in Ukrainian, and their translated English versions belong to publishers from other countries. As a rule, the English versions of these editions are indexed in the international bibliographic databases. Thus, the actual number of published highly cited articles in Ukrainian editions may be somewhat different.

Among the highly cited papers of Ukrainian scientists, the most papers that were written by a large circle of co-authors, collected in this study in the group "Multi5+" (more than 5 co-authors). Second by the level of citedness were documents "Group 3-5" (3-5 co-authors). The result was quite predictable because it was previously established that highly cited articles, as a rule, were written by a large number of scientists from different countries of the world [1]. Therefore, we can state that participation in international collaborations for Ukrainian scientists is a much more effective way of presenting the results of research than individual, or publications in national collaborations.

However, it is also important that from 16% to 27% of the highly cited articles of Ukrainian scientists were written without the participation of foreign partners. It means that Ukraine alone manages to get a significant part of the scientific results visible to the academic community, despite limited funding and miscalculations in the state administration of science. At the same time, it is noteworthy that the Ukrainian authors of the documents of the top 10% in 2011-2015 according to Scopus, are represented both by the institutions of the National Academy of Sciences of Ukraine and by Ukrainian universities, that is, they are engaged not only in scientific work but also conduct educational activities. The experience of successful scientific teams and individual Ukrainian scientists requires a separate analysis, which can be used in the future for developing of a new roadmap of Ukrainian science.

## 6. Appendix

**Table 5.** The Subject Area Categories of journal publications of Ukrainian scientists of the top 1%, 5% and 10% most cited documents in 2011-2015.

| Subject Area | 1% | 5% | 10% |
|---|---|---|---|
| Physics and Astronomy | 97 | 496 | 959 |
| Materials Science | 23 | 181 | 383 |
| Engineering | 33 | 176 | 343 |
| Chemistry | 12 | 99 | 237 |
| Medicine | 59 | 149 | 231 |
| Mathematics | 16 | 85 | 201 |
| Biochemistry, Genetics and Molecular Biology | 24 | 97 | 169 |
| Earth and Planetary Sciences | 9 | 52 | 115 |
| Agricultural and Biological Sciences | 7 | 48 | 96 |
| Chemical Engineering | 8 | 44 | 87 |
| Computer Science | 8 | 36 | 74 |
| Energy | 9 | 47 | 72 |
| Environmental Science | 6 | 29 | 56 |
| Pharmacology, Toxicology and Pharmaceutics | 3 | 26 | 54 |
| Social Sciences | 9 | 31 | 53 |
| Arts and Humanities | 4 | 23 | 31 |
| Immunology and Microbiology | 7 | 19 | 26 |
| Economics, Econometrics and Finance | 2 | 6 | 21 |
| Multidisciplinary | 2 | 12 | 18 |
| Health Professions | 1 | 13 | 17 |
| Neuroscience | 2 | 6 | 15 |
| Psychology | 2 | 6 | 9 |
| Business | 0 | 4 | 8 |
| Management and Accounting | 0 | 4 | 8 |
| Decision Sciences | 1 | 3 | 6 |
| Nursing | 0 | 2 | 5 |
| Dentistry | 0 | 2 | 2 |
| Veterinary | 1 | 1 | 2 |

**Table 6.** Thematic distribution of individual or written only in national collaborations journal publications of Ukrainian scientists of the top 1%, 5% and 10% most cited documents in 2011-2015.

| Subject Area | 1% | 5% | 10% |
|---|---|---|---|
| Physics and Astronomy | 2 | 47 | 111 |
| Engineering | 8 | 51 | 104 |
| Materials Science | 2 | 25 | 74 |
| Mathematics | 2 | 16 | 48 |
| Chemistry | 2 | 14 | 38 |
| Energy | 4 | 25 | 32 |
| Biochemistry, Genetics and Molecular Biology | 2 | 13 | 29 |
| Computer Science | 3 | 9 | 23 |
| Earth and Planetary Sciences | 0 | 6 | 20 |
| Social Sciences | 0 | 7 | 16 |
| Economics, Econometrics and Finance | 1 | 4 | 14 |
| Medicine | 0 | 4 | 14 |
| Pharmacology, Toxicology and Pharmaceutics | 0 | 5 | 11 |
| Arts and Humanities | 0 | 5 | 8 |
| Neuroscience | 0 | 3 | 8 |
| Chemical Engineering | 3 | 4 | 7 |
| Agricultural and Biological Sciences | 2 | 3 | 5 |
| Health Professions | 0 | 3 | 5 |
| Environmental Science | 0 | 2 | 4 |
| Business, Management and Accounting | 0 | 0 | 2 |
| Decision Sciences | 1 | 1 | 2 |
| Immunology and Microbiology | 0 | 0 | 1 |

## 7. References

1. Aksnes, D. W. Characteristics of highly cited papers. *Research Evaluation*, *12* (3), 2003, 159–170. doi:10.3152/147154403781776645

2. Aksnes, D. W. Citation rates and perceptions of scientific contribution. *Journal of the American Society for Information Science and Technology*, *57* (2), 2006, 169–185. doi:10.1002/asi.20262

3. Ferligoj, A., Kronegger, L., Mali, F., Snijders, T. A. B., & Doreian, P. Scientific collaboration dynamics in a national scientific system. *Scientometrics*, *104* (3), 2015, 985–1012. doi:10.1007/s11192-015-1585-7

4. Glanzel, W. National characteristics in international scientific co-authorship relations. *Scientometrics*, *51* (1), 2001, 69–115. doi:10.1023/A:1010512628145

5. Katz, J. S., & Martin, B. R. What is research collaboration? *Research Policy*, *26* (1), 1997, 1–18. doi:10.1016/S0048-7333(96)00917-1

6. Kavunenko, L., Khorevin, V., & Luzan, K. Comparative analysis of journals on social sciences and humanities in Ukraine and the world. *Scientometrics*, *66* (1), 2006, 123–132. doi:10.1007/s11192-006-0009-0

7. Schiermeier, Q. Ukraine joins flagship European research programme. *Nature*, 2015. doi:10.1038/nature.2015.17164

8. Schiermeier, Q. Conflicting laws threaten Ukrainian science. *Nature*, *531* (7592), 2016, 18–19. doi:10.1038/531018a

9. Schubert, A., & Glänzel, W. Cross-national preference in co-authorship, references and citations. *Scientometrics*, *69* (2), 2006, 409–428. doi:10.1007/s11192-006-0160-7

10. Sooryamoorthy, R. Do types of collaboration change citation? Collaboration and citation patterns of South African science publications. *Scientometrics*, *81* (1), 2009, 177–193. doi:10.1007/s11192-009-2126-z

11. Ynalvez, M. A., & Shrum, W. M. Professional networks, scientific collaboration, and publication productivity in resource-constrained research institutions in a developing country. *Research Policy*, *40* (2), 2011, 204–216. doi:10.1016/j.respol.2010.10.004